\documentclass[showpacs,showkeys,twocolumn,amsmath,amssymb]{revtex4}

\usepackage{graphicx,epsfig}
\usepackage{dcolumn}
\usepackage{bm}
\newcommand{\pr}{\text{Pr}}

\newcommand{\tp}{\tilde{p}}

\newcommand{\ie}{\it i.e. \rm}
\newcommand{\eg}{\it e.g. \rm }

\begin{document}

\preprint{}

\title{Sampling properties of random graphs: the degree distribution}
\author{Michael P.H. Stumpf}
\email{m.stumpf@imperial.ac.uk}
 \homepage{http://www.bio.ic.ac.uk/research/stumpf}
\affiliation{Centre for Bioinformatics, Division of Molecular Biosciences, Imperial College London, Wolfson Building, London SW7 2AZ, UK}
\author{Carsten Wiuf}
\email{wiuf@birc.au.dk}
 \homepage{http://www.birc.au.dk/~wiuf}
\affiliation{Bioinformatics Research Center, University of Aarhus, 8000 Aarhus C, Denmark}

\date{\today}

\begin{abstract}
We discuss two sampling schemes for selecting random subnets from a network: Random
sampling and connectivity dependent sampling, and investigate how the degree
distribution of a node in the network is affected by the two types of sampling. Here
we derive a necessary and sufficient condition that guarantees that the degree distribution of the subnet and the true network belong to the same family of probability distributions.
For completely random sampling of nodes we find that this condition is fulfilled by classical random graphs; for the vast majority of networks this condition will, however, not be met. We  furthermore discuss the case where the probability of sampling a node depends on the degree of a node and we find that even classical random graphs are no longer closed under this sampling regime. We conclude by relating the results to real {\it E.coli} protein interaction network data.
\end{abstract}

\pacs{02.50.-r, 89.75.Hc, 89.75.Fb}
\keywords{Complex networks, systems biology, protein interaction networks}
\maketitle

\section{\label{sec:level1a}Introduction}

Most networks investigated today are parts of much larger networks. These subnets can come in two different forms: first, we can choose a region of a network and consider all nodes that are in this region and only the edges between these nodes (for example a connected component of the larger network
would be one such subnet). Looking at networks  defined by all servers in a country, or the interaction network of all proteins which are confined to the mitochondria would be real-world examples\cite{Albert2002,Newman2003b,Evans2004}. Such networks may not be representative of the network as a whole but can give valuable insights into communication or biological processes within a defined sphere. More complicated is a second type of subnet where each node of the global network is included in the subnet with a certain probability $p$ and only the connections between pairs of nodes which are both included in the subnet are studied. This type of subnet is radically different from the regional-based subnets. It is, however, a frequent scenario in the analysis of technological and biological networks: most studies of molecular networks, such as protein-protein interaction\cite{Maslov2002,Agrafioti2005}, gene-regulation\cite{Xiong2004} and metabolic networks\cite{Bhattacharya2003}, test for connections between a subset of the known molecular entities (proteins, genes and enzymes/metabolites, respectively). The process by which these entities (or corresponding probes) are chosen may reflect the bias of the experimenter or merely chance, and this will in turn influence the extent to which the subnet reflects properties of the global network in a meaningful way. In light of the relative straightforwardness of studying the sampling properties of networks, and their obvious importance for the analysis of current network data sets it is surprising that this problem has not been addressed previously. 
\par
\begin{figure}
\epsfig{file=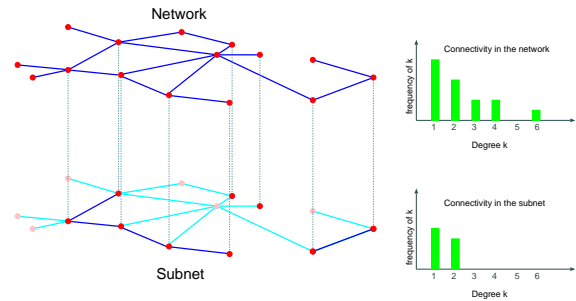,width=7.5cm}
\caption{Sampling nodes from the network (top) will give rise to subnets (bottom). If edges are only observed if both nodes incident on an edge are included in the subnet (indicated in dark blue), then the degree distributions (as well as other characteristics) of the subnet and global network will be different. In the text we show that sometimes, however, degree distributions in both networks can be related under random sampling of nodes.} 
\label{fig0}
\end{figure}
Here we will focus on the simplest, and perhaps
most parsimonious, process of sampling nodes: the case where each node in the network is included with probability $0<p<1$. In the present analysis we will concentrate on the sampling properties of the degree distribution of a network. The degree distribution, henceforth denoted by $\pr(k)$, specifies the probability for a node to have $k$ connections, $k=0,1,\ldots$, and is probably the most common summary statistic used in the analysis of networks \cite{Stumpf2005a}. In particular the potential scale-free nature of real networks is often identified from the empirical degree distribution, which for scale-free networks takes on a power-law form, $\pr(k)\propto k^{-\gamma}$ \cite{Bollobas2003,Newman2003b,Dorogovtsev2003,Stumpf2005b}. Frequently a model is considered scale-free if the tail (\ie for $k$ sufficiently large) of the degree distribution takes such an asymptotic powerlaw form \cite{Barabasi1999b,Albert2002}. Here we will consider this case as well as network ensembles with an exact power-law degree distribution.  The central question addressed here is whether the degree distribution of randomly sampled subnets has the same properties as the degree distribution of the overall network. Thus far this question has been ignored in the literature, but as we will show, is of great importance in the analysis of real networks, which in their vast majority, are only subnets of larger networks. Unless explicitly stated otherwise we shall consider the thermodynamic limit, $N\rightarrow \infty$.
\par

\section{\label{sec:level1b}The degree distribution of a random subnet}
\subsection{\label{sec:level2b}Sampling from networks}
We use $\cal{N}$ to denote a network with $N$ nodes (we allow $N\rightarrow \infty$) drawn from a statistical ensemble of random networks \cite{Bollobas1998,Burda2001} defined by some (potentially vector valued) parameter $\Omega$ and let $\pr(k)$ be its degree distribution; the total number of edges is given by $M$. Here we will be especially concerned with the case of a subnet $\cal{S}$ generated from the global network $\cal{N}$ by randomly sampling each node $i\in \cal{N}$ with a certain probability $0\le p_i\le 1$. Thus if a node of degree $k$ gets picked for inclusion in the subnet, its degree in the subnet will depend on the number of its neighbours which are also included in $\cal{S}$.

\subsubsection{\label{sec:level3} Random sampling}
We start by considering the case where the probability of picking a node is identical for all nodes, $p_i=p$ for all $i$. Here $p=0$ and $p=1$ are the trivial cases for which $\cal{S}=\emptyset$ and $\cal{S}=\cal{N}$, respectively.
Formally, the probability that a node has connectivity $l$ in $\cal{S}$ given it has connectivity $k$ in $\cal{N}$ is
\begin{equation}
\pr(l|k)= \binom{k}{l} p^l(1-p)^{k-l},
\label{ptrans}
\end{equation}
where $\pr(x|y)$ denotes the conditional probability of $x$ given $y$. The degree distribution in the subnet is thus given by
\begin{equation}
\pr_{\cal S}(l) = \sum_{k\ge l}^\infty \pr(l|k)\pr(k) = \sum_{k\ge l}^\infty  \binom{k}{l} p^l(1-p)^{k-l}\pr(k).
\label{prsubnet}
\end{equation}
This is probably the simplest and most parsimonious sampling scheme and may also be a reasonably realistic approximation, \eg in the study of protein interaction networks where experimenters choose a set of proteins in a more or less haphazard fashion. 
\par
From Eqn. (\ref{prsubnet}) we can show that
\begin{equation}
\Bbb{E}_{\cal S}[l]=\Bbb{E}\left[\Bbb{E}[l|k]\right] = p \,\Bbb{E}[k] = p\tau,
\label{means}
\end{equation}
where $\tau := \Bbb{E}[k]$ is the average degree in the network.
 Similarly we can show that the $m$-th moment of the descending factorial (defined by $x_{[m]}=x(x-1)(x-2)\dots (x-m+1)$ \cite{Feller1957}) for the degree distribution of a network obeys
\begin{equation}
\Bbb{E}_{\cal S}[l_{[m]}] = p^m \Bbb{E}[k_{[m]}].
\label{moments}
\end{equation}
Eqns. (\ref{means}) and (\ref{moments}) are fulfilled for all networks, as long as the moments exist; for scale-free networks with exponent $\gamma$, for example, moments of order greater or equal than $\lfloor \gamma \rfloor$ do not exist.

\subsubsection{Random sampling dependent on degree}
A further sampling scheme will be considered here where the number of connections directly influences the probability, $\pi(k)$, of sampling a node of degree $k$; In the previous sampling scheme all nodes had the same chance of being sampled, $\pi(k)=p$. We will focus on the particular case of an uncorrelated network.
\par
The connectivity of a node in the subnet thus depends on the degrees of its neighbours. 
The probability that a node connected to a randomly chosen edge has degree $k$ is given by
\begin{equation}
\pr^*(k) = \frac{k \pr(k)}{\tau} 
\end{equation}
where $\tau$ is the average degree in the network; the average degree of the neighbours of a randomly chosen node is thus $\Bbb{E}[k^2]/\Bbb{E}[k]$, if the two first moments of the degree distribution exist; below we will limit ourselves to such situations (for finite networks the moments will, of course, exist). Assuming a node is retained in the subnet then the probability of sampling a neighbouring node depends also on its connectivity and, in a mean-field approximation, the probability of retaining an edge originating from a node, $\tilde{p}$ is thus given by
\begin{equation}
\tp = \frac{1}{\tau}\sum_k k\pr(k)\pi(k)
\label{tilde}
\end{equation}
The degree distribution of the subnet $\cal S$  is again given by  binomial sampling:
\begin{equation}
\pr_{\cal S}(l)\!= \!\left[\sum_{k\ge l}^\infty \binom{k}{l} \tilde{p}^l(1-\tilde{p})^{k-l}\pi(k)\pr(k)\right]/\sum_{k=0}^\infty \pi(k)\pr(k).
\label{prconnect}
\end{equation}
Defining 
\begin{equation}
\pr_0(k)= \pi(k)\pr(k)/\sum_{k=0}^{\infty} \pi(k)\pr(k) 
\label{prdef}
\end{equation}
we can rewrite Eqn. (\ref{prconnect}) in the same form as Eqn. (\ref{prsubnet}).
With these probabilities the degree distribution in the subnet is given analogously to Eqn. (\ref{prsubnet}) as 
\begin{equation}
\pr_{\cal S}(l) = \sum_{k\ge l}^\infty \binom{k}{l} \tilde{p}^l(1-\tilde{p})^{k-l}\pr_0(k)
\label{prconnet2}
\end{equation}
Obviously, when setting $\pi(k)=p$ Eqn. (\ref{prconnet2}) simplifies to Eqn. (\ref{prsubnet}). 
\par
We still have to specify the functional form of $\pi(k)$; {\em a priori} the only constraint is that $\pi(k)$ has to be a probability for all $k$, \ie $0\le \pi(k)\le 1, \forall k = 0, 1, 2 \ldots$. One possible and obvious choice is to let $\pi(k)\propto k$; in order to ensure that $\pi(k)\le 1$ for large $k$ we set
\begin{equation}
\pi(k) = Ck
\label{eqdefpi}
\end{equation}
with $C$ sufficiently small such that $\pi<1$ for large $k$ (we can always trivially set $C=2\Bbb{E}[M])$ with $\Bbb{E}[M]$ the expected number of edges in the network). In this case
\begin{equation}
\tp = \frac{C}{\tau}\sum_kk^2 \pr(k) = C\Bbb{E}[k^2],
\label{propsample}
\end{equation}
\ie $\tp$ depends on the degree distribution solely via the first and second moments of $\pr(k)$.
 We will refer to this sampling scheme as preferential sampling of nodes.

\subsection{\label{sec:level2c}Probability generating functions of random subnets}
We represent the degree distribution of a network $\cal{N}$ through its probability generating function (PGF) \cite{Feller1957,Newman2003e},
\begin{equation}
G(s) = \sum_{k=0}^\infty \pr(k) s^k.
\label{pgfnet}
\end{equation}
The probability $\pr(k)$ follows from the PGF via the relationship
\begin{equation}
\Pr(k) = \left. \frac{1}{k!}\frac{d^k G(s)}{d s^k}\right|_{s=0}
\end{equation}
\par
With Eqns. (\ref{pgfnet}) and (\ref{ptrans}) we can straightforwardly derive the PGF for the subnet
\begin{align}
G_{\cal{S}}(s)& =\sum_{l=0}^\infty \pr(l) s^l\nonumber\\
&= \sum_{l=0}^\infty s^l\sum_{k=l}^\infty \pr(l|k)\pr(k)\nonumber\\
&=\sum_{k=0}^\infty\sum_{l=0}^k\binom{k}{l}p^l(1-p)^{k-l}s^l\pr(k)\nonumber\\
&=\sum_{k=0}^\infty \pr(k)(1-p+ps)^k\nonumber\\
&=G(1-p+ps).
\label{subnetpgf0}
\end{align}
If nodes with degree $l=0$ are ignored (as is frequently the case in high throughput protein interaction data) then after deleting all nodes with $l=0$ the PGF in the subnet becomes
\begin{align}
G_{\cal{S}}^*(s)=\frac{G(1-p+ps)-G(1-p)}{1-G(1-p)}.
\label{subnetpgf1}
\end{align}
Eqns. (\ref{subnetpgf0}) and (\ref{subnetpgf1}), respectively, hold  generally for the degree distributions of subnets randomly sampled from networks, depending on whether orphaned nodes (\ie those with connectivity $l=0$) are allowed or not \cite{Stumpf2005}.
\par
Interestingly, if Eqn. (\ref{subnetpgf0}) holds then also Eqn. (\ref{subnetpgf1}) holds with
$G(s)$ replaced by $G^*(s)=(G(s)-\pr(0))/(1-\pr(0))$; \ie networks with orphaned nodes
removed are closed under random sampling if the networks with the orphaned nodes retained are.
\par

\section{\label{sec:level1c} Closure under random sampling from networks}
\subsection{Conditions for closure: generating function}
From Eqns. (\ref{subnetpgf0}) and (\ref{subnetpgf1}) it is apparent that degree distributions of  a subnet $\cal{S}$ cannot generally be expected to be of the same type 
({\it e.g.} a Possion distribution) as the degree distribution of the global network $\cal{N}$. For some important types of networks, however, it can be shown that random sampling of nodes gives rise to networks with  degree distributions  of the same type as the global network, but with a different parameter depending on $p$, {\it i.e.} $\Omega'=fn(\Omega,p)$. In this case we say that a network (or its degree distribution) is closed under random sampling of nodes. For a network ensemble to be closed under random sampling the following condition is necessary and sufficient \cite{Stumpf2005},
\begin{equation}
G_{\cal{S}}(s;\Omega) = G(s;\Omega')=G(1-p+ps;\Omega),
\label{closure1}
\end{equation}
 and 
\begin{equation}
G_{\cal{S}}^*(s;\Omega) = G(s;\Omega')=\frac{G(1-p+ps;\Omega)-G(1-p;\Omega)}{1-G(1-p;\Omega)},
\label{closure2}
\end{equation}
when the subnet is not allowed to have orphaned nodes. Necessity and sufficiency follow from Eqns. (\ref{subnetpgf0}) and (\ref{subnetpgf1}) and the definition of the properties of a closed subnet.

\subsection{Conditions for closure: moments}
Equations (\ref{closure1}) and (\ref{closure2}) can be applied to all types of degree distributions. Inspired by  Eqns. (\ref{means}) and (\ref{moments}) we  here derive a general condition in terms of moments for a subnet to be of the same type as the global network. We assume the moments determine the degree distribution uniquely (in particular, this implies that all moments exist), which is true under mild regularity conditions.
Let an ensemble of random networks be given which is parameterized by $\Omega$. 
For example, the ensemble of classical or Erd\"os-R\'enyi random graphs \cite{Erdos1959,Bollobas1998} has $\pr(k)=\exp(-\lambda)\frac{\lambda^k}{k!}$  and $\Omega=\lambda$ is the average connectivity. We seek a condition that, provided nodes are sampled with probability $p$, ensures that the degree distribution of the subnet remains in the same ensemble of random networks. Without loss of generality we can assume that $\Omega$ has the form $\Omega=(\tau,\psi)$, where $\tau$ is the average degree in the network and $\psi$ is an additional (potentially vector valued) parameter.
\par
From Eqn. (\ref{means}) we know that the average connectivity in the sampled subnet, $\tau_p$, is given by $\tau_p=p\tau$.
We can use Eqn. (\ref{moments}) to show that a family of degree distributions is closed under random sampling of nodes if and only if the descending factorial moments obey the relationship
\begin{equation}
\Bbb{E}[k_{[m]}] = a_m(\psi) \tau^m,
\label{relationship}
\end{equation}
where $a_m(\psi)$ is a constant that depends only on $m$ and $\psi$ but not on $\tau$ and the sampling probability $p$, and where $a_1(\psi)=1$. 
\par
To prove that Eqn. (\ref{relationship}) is necessary we assume that the network is closed under random sampling of nodes and write $\tau=\Bbb{E}(k)$ and $g_m(\tau,\psi)=\Bbb{E}(k_{[m]})$. Because of Eqns. (\ref{means}) and (\ref{moments}) we can immediately write
\begin{equation}
g_m(p\tau,\psi) = p^m g_m(\tau,\psi)\\
\end{equation}
and
\begin{equation}
\frac{g_m(p\tau,\psi)}{(p\tau)^m}= \frac{g_m(\tau,\psi)}{\tau^m}.
\end{equation}
Thus $g_m(\tau,\psi)/\tau^m = \text{const.}$ (for all $\tau$) or
\begin{equation}
g_m(\tau,\psi) = a_m(\psi)\tau^m, 
\end{equation}
with $a_1(\psi)=1$ as required. 
\par
To prove sufficiency assume that the descending moments of $k_{[m]}$ fulfil Eqn. (\ref{relationship}); using Eqn. (\ref{moments}) the descending factorial moments of the nodal degrees in the subnet follow the relationship
\begin{equation}
\Bbb{E}_{\cal S}[l_{[m]}] = a_m(\psi)(p\tau)^m.
\end{equation}
Since the descending moments determine the moments, $\Bbb{E}(k^m)$ of a degree distribution, which in turn determine the distribution uniquely (by assumption), then the degree distribution of the subnet is given by a distribution that is of the same type as the degree distribution but with a rescaled parameter. Thus Eqn. (\ref{relationship}) is a necessary and sufficient condition for a network ensemble to be closed under random sampling of nodes.  $\Box$

\subsection{Analytical Examples}
We can use relationships (\ref{closure1}) and (\ref{relationship}) to determine whether a degree distribution is closed under random sampling. We will discuss this for three commonly observed degree distributions. Note that we only consider a degree distribution to be closed under (random) sampling if the degree distributions of the network and the subnet belong to the same family of probability distributions.
\par
Classical random graphs have a Poisson degree distribution, $Po(\lambda)$. It is straightforward to show that the descending moments of the Poisson distributed random variables are given by
\begin{equation}
\Bbb{E}[k_{[m]}] = \tau^m = \lambda^m.
\end{equation}
Thus $a_m=1$ for all 
$m\ge 1$ and the degree distribution of classical random graphs is closed under random sampling of nodes. If we therefore have a subnet $\cal{S}$ of size $M$ drawn from a larger network $\cal{N}$ of known size $N$ we can determine $\lambda$ from $\lambda_{\cal{S}}$ as $\lambda=\lambda_{\cal{S}}\frac{N}{M}$. The subnet is therefore informative about the global network.
\par
Networks which grow by random attachment of new nodes give rise to exponential degree distributions such that asymptotically (large $N$) $\pr(k) = (1-e^{-\alpha})e^{-k\alpha}$. For such a distribution it is easily shown that  
\begin{equation}
\Bbb{E}[k_{[m]}] = \frac{m! e^{-m\alpha}}{(1-e^{-\alpha})^m} = m!\tau^m,
\end{equation}
since $\Bbb{E}[k] = e^{-\alpha}/(1-e^{-\alpha})$. This means that $\Bbb{E}[k_{[m]}]$ can be written in the form specified by Eqn. (\ref{relationship}) and therefore exponential degree distributions are closed under random sampling. Binomial (as for classical finite-sized random graphs) and negative binomial distributions are also closed under random sampling as is easily verified. An explicit construction of probability distributions which are closed is discussed in appendix A.
\par
If the probability of attaching to a node is proportional to its degree the resulting network will asymptotically have a power-law degree distribution with exponent 3 \cite{Barabasi1999b}. For models where an existing  node is duplicated and each of its connections is kept with certain probability degree distributions will also be power-law like but with exponents $2<\gamma<3$ \cite{Aiello2001}.
\par
We first consider the sampling properties of network ensembles with degree distribution given by an exact powerlaw, $\pr(k)= k^{-\gamma}/\zeta(\gamma)$. In the asymptotic limit, $N\rightarrow \infty$, all moments greater than $\lceil \gamma\rceil$ diverge and we therefore have to use the PGF formalism. The PGF for the global network is given by
\begin{equation}
G(s;\gamma) = \frac{1}{\zeta(\gamma)}\sum_{k=1}^\infty s^k k^{-\gamma}
\end{equation}
and since $k=0$ is explicitely forbidden in a scale-free network, we use Eqn. (\ref{closure2}) to construct the PGF in the subnet, whence 
\begin{equation}
G_{\cal{S}}^*(s;\gamma) =\frac{\sum_{k=1}^\infty \left[(1-p+ps)^k-(1-p)^k\right]k^{-\gamma}}{\zeta(\gamma)-\sum_{k=1}^\infty (1-p)^kk^{-\gamma}}.
\end{equation}
Clearly for $p\rightarrow 1$ we obtain the original PGF, $G(s;\gamma)$. For $0<p<1$, however, it is impossible to determine an exponent $\gamma'$ such that $G_{\cal S}$ could be written in terms of the PGF of a power law. Therefore random subnets drawn from exact scale-free networks are not themselves scale-free. This can also be shown explicitely using a series expansion \cite{Stumpf2005}. We note, however that the tail of the degree distribution of the subnet still takes on a powerlaw form for $k$ sufficiently large. 
The same analysis applied to other fat-tailed probability distributions also shows that other fat-tailed degree distributions such as the log-normal and the stretched exponential families \cite{Sornette2003} are not closed under random sampling.
\par

\subsection{Numerical Examples}
The effect of random sampling on the degree distribution is most straightforwardly illustrated using numerical solutions of Eqns. (\ref{prsubnet}) and (\ref{tilde}-\ref{prconnet2}.) Here we do this for networks of infinite size and for simplicity focus on the canonical models of the classical random graph and the exact scale-free network, respectively. 
\par
\begin{figure}
\epsfig{file=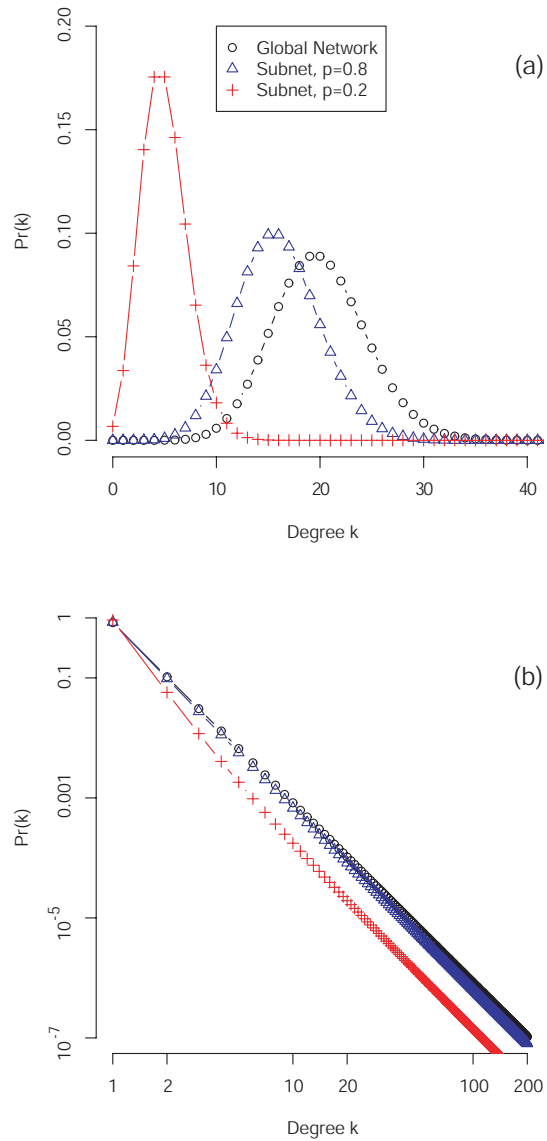,width=8cm}
\caption{Degree distributions of full network and subnets obtained by sampling each node with probability $p=0.8$ and $p=0.2$, respectively, for classical random graphs (a) and scale-free networks (b).}
\label{fig1}
\end{figure}
In part (a) of figure \ref{fig1} we show the Poisson distribution with $\lambda=5$ and the distributions of random subnet with $p=0.8$ and $p=0.2$, respectively. The subnet distributions are identical with the Poisson distributions with parameters $\lambda=4$ and $\lambda=2$. This also means that as $p\lambda$ becomes smaller than one the subnet will move through the phase-transition where the giant connected component dissolves and the size distribution of connected parts of the subnet becomes exponential.
\par
In part (b) of the same figure we show the power-law distribution with $\gamma=3$ and again the respective subnet degree distributions (renormalized such that $\pr_{\cal S}(0)=0$ in the subnet). We find that the subnet degree distributions are no longer straight lines but that as $k$ becomes large they run parallel to the original distributions. That is, as already described above, the tails of degree distributions of subnets sampled randomly from scale-free networks also fall off in the same power-law fashion as the original network. But at low connectivities the departure from the scale-free network is quite pronounced: probability mass moves from the tail towards the lowly connected nodes with $k=1$, which become more abundant than would be expected for a true scale-free network. This will have quite considerable effects for finite size networks. The deviation of the subnet degree distribution from a pure power-law at small to intermediate connectivities increases with $\gamma$ (as well as, naturally, with decreasing sampling probability $p$). We note however, that the tail of the degree distribution will retain a powerlaw form; thus for an alternative definition of scale-free behaviour which only requires $\pr_{\cal N}(k)\propto k^{-\gamma}$ for $k\rightarrow \infty$ random subnets will retain scale-free behaviour in the sense that the tail is still described by a powerlaw $\pr_{\cal S}(k)\propto k^{-\gamma'}$ for $k\rightarrow \infty$. In general, however, when the whole degree distribution is considered scale-free networks are not closed under random sampling.

\section{Connectivity dependent sampling}
There is no unique and obvious way in which the probability of sampling a node may depend on the connectivity. Here we briefly outline the behaviour of the degree distribution under the simple schemes outlined above where the probability of sampling a node is no longer uniform but linearly proportional to its connectivity, \ie if $\pi(k)\propto k$; we assume that $\tp(k)$ is given by Eqn. (\ref{propsample}). 
\par
For a Poisson degree distribution with parameter $\lambda$ we have $\Bbb{E}[k^2]= \tau^2+\tau=\lambda^2+\lambda$ and 
$\Bbb{E}[M] =N\lambda/2$, (assuming the network is large and finite)
whence $\tp=(\lambda+1)/(N\lambda)$ and
\begin{equation}
\pr_0(k)= \frac{e^{-\lambda}\lambda^{k-1}}{(k-1)!},
\end{equation}
if we set $C=2\Bbb{E}[M]$ in Eqn. (\ref{eqdefpi}). 
In this case Eqn. (\ref{prconnet2}) becomes
\begin{align}
\pr_{\cal{S}}(l) = &\sum_{k\geq l}^{\infty}\binom{k}{l}\tilde{p}^l(1-\tilde{p})^{k-l}\pr_0(k)\nonumber\\
&= \frac{(\lambda\tilde{p})^l}{l!}e^{-\lambda\tilde{p}}\left(1-\tilde{p}+\frac{l}{\lambda}\right)
\label{prersample}
\end{align}
for $l=0,1,\ldots$.
The distribution in the subnet is thus not a pure Poisson distribution but one multiplied by a factor $1-\tilde{p}+l/\lambda$. Under this connectivity dependent sampling classical random graphs are therefore not closed and subnets $\cal S$ are qualitatively (if perhaps only rather slightly) different from the overall network $\cal N$.
\par
For scale-free networks with $\gamma\le 3$ the second moment diverges, $\Bbb{E}[k^2]\rightarrow \infty$, and we therefore focus on finite (though potentially very large networks). Networks with a powerlaw degree distribution can, for example, be constructed using standard methods \cite{Bender1978,Molloy1995,Newman2001a}. For such a scale-free graph with $N$ nodes we have to numerically evaluate the expected number of edges $\Bbb{E}[M]=\frac{1}{2}\sum_{k=1}^N k^{-\gamma+1}/\zeta(\gamma)$ and $\tp$, given by Eqn. (\ref{propsample}). For $\pr_0(k)$ we obtain for scale-free networks
\begin{equation}
\pr_0(k) = \frac{k^{1-\gamma}}{\zeta(\gamma-1)}. 
\end{equation}
Proportional sampling from a scale-free network defined by a powerlaw exponent $\gamma$ is thus identical to sampling from a network with powerlaw exponent $\gamma-1$ and sampling probability $\tp$. Therefore we can use the results obtained above and conclude that the scale-free network (in the strict sense outlined above) is not closed under proportional sapling of nodes; for sufficiently large degrees, however, the tail of the degree distribution will still have a powerlaw form.

\section{Protein Interaction Network Data}
\begin{figure}
\epsfig{file=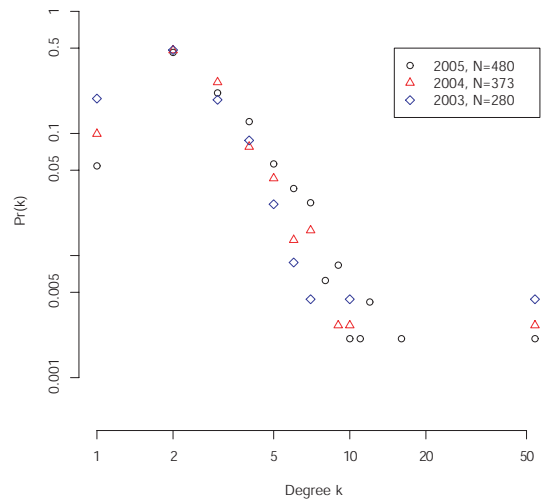,width=8cm}
\caption{Degree distributions of protein interaction network data available for {\em E.coli} in April of the years 2003, 2004 and 2005, respectively. As the fraction of sampled nodes/proteins decreases statistical weight is shifted from the tail towards lower degrees.}
\label{fig3}
\end{figure}
In figure \ref{fig3} we show three degree distributions corresponding to the protein interaction network (PIN) data from {\em E.coli} which was available in April 2003, 2004 and 2005 in the database of interaction proteins (DIP; \url{dip.doe-mbi.ucla.edu}); the resulting networks are made up of the interactions among 228, 373 and 480 proteins and have 293, 515 and 760 interactions, respectively. Figure \ref{fig3} confirms the results of the theoretical analysis presented above: as the fraction of sampled network nodes decreases statistical weight shifts from the tail towards lower degrees; the degree of the single highly connected node, $k=54$, was already known in the 2003 dataset (no further interactions have been added to this node since). The statistical weight of sparsely connected nodes, $k=1$, increases as the fraction of sampled nodes decreases. We note that the present data samples only a small subnet of the {\em E.coli} PIN which consists of interactions among approximately 4000 proteins. Moreover (i) it is well established that PIN data is highly unreliable and very noisy, and (ii) the true sampling scheme underlying the sampling scheme will generally be more complicated than the first order model employed here. The behaviour appears, however, to be qualitatively similar to our theoretical analysis.

\section{Conclusion}
Both sampling schemes discussed here are necessarily simpler than is the case in many real situations, such as the analysis of protein interaction networks (see \eg \cite{Ito2000,Mering2002}. We believe, however, that between them they retain some vestiges of reality. Crucially, however, we wish to stress the incomplete nature of many network data sets. For many of these data sets in fact, including protein interaction network data, it appears that some form of random sampling is more realistic than a process in which the neighbourhood of a node is explored and neighbouring sites are recruited iteratively into the experimental setup. No matter what the sampling process is, it has to be included into the analysis from the outset: making inferences from incomplete (in the sense that not all nodes have been sampled) network data may give misleading results. If a network is closed under random (or connectivity dependent)sampling then it is straightforward to infer properties of the overall network from the subnet. For some, notably Erd\"os-R\'enyi random graphs, this is indeed the case. In general, however, the degree distributions of the network and sampled subnets will be qualitatively different. For example, while powerlaw tails will also give rise to powerlaw tails in the subnet a network which has an exact powerlaw degree distribution is not closed under random sampling. The same is true for other broad-tailed degree distributions such as lognormal or stretched exponential distributions.
\par
Sampling properties will also affect other network statistics, including network diameter and average path length, clustering coefficient and network motifs. These will be studied in a companion paper. We believe that sampling properties ought to be included explicitly and from the outset into any network analysis, unless there is good evidence that the whole (or the majority) of the network's nodes have been included in the data. Quite apart from the relevance of this work in the analysis of real data we believe that a detailed analysis of sampling properties of graphs is a rich field which, surprisingly, appears to have been neglected thus far.

\section{Acknowledgement}
MPHS is a Wellcome Trust research fellow and CW is supported by the Danish Cancer Society. Financial support from the Royal Society and the Carlsberg Foundation (to MPHS and CW) is also gratefully acknowledged. We also thank R.M. May, E. de Silva, M.J.E. Sternberg, M. Howard, P.J. Ingram and H.F. Jensen for many useful discussions and helpful comments. The manuscript benefitted greatly from an anonymous referee's comments.
\par
\bigskip

\appendix
\section{Construction of closed degree distributions}
We have shown that Eqn. (\ref{relationship}) is both a necessary and sufficient condition for a degree distribution to be closed under binomial random sampling. We can also use Eqn. (\ref{relationship}) to construct closed distributions {\it de novo} as any series of positive numbers $a_k$, $k=1,2,\ldots$ with $a_1=1$ defines a family of random variables  closed under binomial sampling via the condition
\begin{equation}
\Bbb{E}[k_{[m]}] = a_m \tau^m
\label{definedist}
\end{equation}
for some $\tau\in T=[0,t]$ and $t\geq 0$. 
\par
First, the degenerate distribution $\pr(k=0) =1$ is defined by $\Bbb{E}[k_{[m]}]=0$ for all $m>0$. Therefore $0$ must be in the interval $T$ and $T$ is non-empty as the degenerate distribution is trivially closed under binomial sampling. Now assume that $\tau\geq 0$ defines the distribution of $k$ through Eqn. (\ref{definedist}). 
Any $\tau^*$ with $0\le \tau^*\le \tau$  defines the degree distribution after binomial sampling of nodes from $k$ with probability $p=\tau^*/\tau$ which, by construction, has  degree distribution given by $\Bbb{E}[l_{[m]}] = a_m(\tau^*)^m$. The distributions defined by Eqn. (\ref{definedist}) are
 therefore closed under random sampling of nodes.
\par Eqn. (\ref{definedist}) can be used to to construct arbitrary degree distributions which are closed under binomial sampling. Nontrivial examples are possible; for example
\begin{equation}
a_k = (k+1)2^{-k} \ \ \ \ \ \ \text{ for } k =1,2,\ldots
\end{equation}
defines a distribution closed under random sampling,
\begin{equation}
\pr(k) = \frac{(2\tau)^k}{k!}(k+1-2\tau)e^{-2\tau}
\label{construc1}
\end{equation}
where $\tau=\Bbb{E}[k]\in[0,0.5]$ (note that for $\tau=0.5$, $\pr(k-1)$ defined by Eqn. (\ref{construc1})is Poisson distributed). 
\bibliography{/home/michael/bibliography/mstbibnet.bib}

\end{document}